# THE IMPACT OF COVID-19, CLIMATE CHANGE AND WORKING FROM HOME ON TRAVEL CHOICES

Authors: Torran Semple, Achille Fonzone and Grigorios Fountas

Transport Research Institute, Edinburgh Napier University, Colinton Road, Edinburgh, EH10 4DT

https://blogs.napier.ac.uk/tri/

**THIS RESEARCH WAS FUNDED BY THE SCOTTISH FUNDING COUNCIL**

**Abstract**

The COVID-19 pandemic changed various aspects of our daily lives, including the way in which we travel and commute. Working from home has become significantly more popular since the beginning of the pandemic and is expected to remain a reality for many people even when COVID-19 no longer poses a threat. Decreased commuting trips may also have environmental benefits as people will be able to reduce their overall travel. Working from home may present an opportunity to accelerate the Scottish Government's 'Mission Zero for transport', which aims to decarbonise the transport sector by 2045, however, meeting this target also depends on restoring faith in public transport, which saw significant decreases in usage during the pandemic, and increasing other forms of sustainable travel (e.g., walking and cycling).

In this study we investigate various aspects of Scottish residents' climate change and COVID-19 perceptions using survey data (n=1,050) collected in Scotland during January 2022. Quota restraints were enforced for age and gender to ensure the survey sample was approximately representative of the Scottish population. The survey also included a discrete choice experiment to investigate mode preferences for commuting trips in different working from home and COVID-19 risk scenarios.

Our findings show that some people still need to be convinced that their travel choices can have an effect on climate change and the spread of infectious diseases. The discrete choice experiment showed that the use of cars for commuting is relatively consistent regardless of the working from home situation, however, bicycles become more popular in high working from home scenarios. As expected, the attractiveness of public transport decreases with increased COVID-19 risk and private modes become more popular.

## TABLE OF CONTENTS

# LISTS OF FIGURES, TABLES, AND ACRONYMS

## LIST OF FIGURES

## LIST OF TABLES

# LIST OF ACRONYMS

*WFH – working from home*

*AT – active travel*

*PT – public transport*

*EV – electric vehicle*

*NHS – National Health Service*

# 1. INTRODUCTION & BACKGROUND

## 1.1. CONTEXT

The Scottish Government's 'Mission Zero for transport' committed to decarbonising the transport network by 2045 (Transport Scotland, 2021). Meeting this target depends on the restoration of public faith in public transport, which saw significant decreases in usage during the height of the pandemic, and also on the promotion of other forms of sustainable travel, e.g., active travel (walking and cycling). This may however be counteracted by the current publicity of the climate crisis, which may foster a shift from cars to more environmentally friendly modes.

The COVID-19 pandemic has caused unprecedented behavioural shifts around world, including a significant increase in working from home (WFH) that is likely to continue in the future to a certain extent, even when COVID-19 no longer poses a public health risk. In addition, there have been various iterations of COVID-19 travel advice since the beginning of the pandemic, which have altered travel perceptions, attitudes, and choices. In particular, the attractiveness of public transport has decreased, whereas more people walked or cycled. The decrease in public transport use is expected to outlast the pandemic (Downey, et al., 2021).

Before COVID-19, commuting accounted for almost one quarter of journeys in Scotland, with two thirds of these trips made by car (Transport Scotland, 2019a). A considerable modal shift in commuting trips, away from cars and towards public transport and active travel, would constitute a significant contribution to achieving the Government's decarbonisation targets and is also likely to prompt changes in mode choices for other trip purposes.

In this research, we explore various aspects of Scottish residents' positions regarding climate change and air-borne infectious diseases, and we investigate how such positions may impact preferred mode of travel choices for commuting trips under different WFH and COVID-19 infection risk scenarios.

## 1.2. RESEARCH CONTENT

In our study, we

- Investigate various aspects of Scottish residents' climate change and COVID-19 perceptions, including attitudes, behaviours, values, social factors, affective factors, and flexibility of habits
- Conduct a discrete choice experiment (DCEs) to determine Scottish residents' commuting preferences (i.e., preferred mode of travel for commuting trips) in four distinct scenarios: (I) WFH occasionally, no COVID-19 risk; (II) WFH occasionally, medium COVID-19 risk; (III) WFH frequently, no COVID-19 risk; and (IV) WFH frequently, medium COVID-19 risk

## 2. RESEARCH APPROACH

### 2.1. DATA COLLECTION

A survey of Scottish residents (n=1,050) was conducted in January 2022 to gauge climate change and COVID-19 perceptions, mobility patterns, and commuting preferences in different WFH and COVID-19 infection risk scenarios. The survey was conducted using the online survey platform Qualtrics. Quota restraints were imposed for gender and age to ensure the survey sample accurately represented the Scottish population (see Section 3.1. for further information on sample representativeness).

### 2.2. THEORETICAL FRAMEWORK

The theoretical framework underpinning the survey design was based on Triandis' theory of interpersonal behaviour (shown in **Figure 1**), which theorises that intention, habit and context are the main determinants of behaviour (Domarchi, et al., 2008). A range of factors affect intention, namely, attitudes (e.g., expected outcomes and value of adopting a given behaviour), social factors (e.g., social norms and social responsibility affecting the perceptions of a given behaviour) and affective factors, which include emotions and feelings.

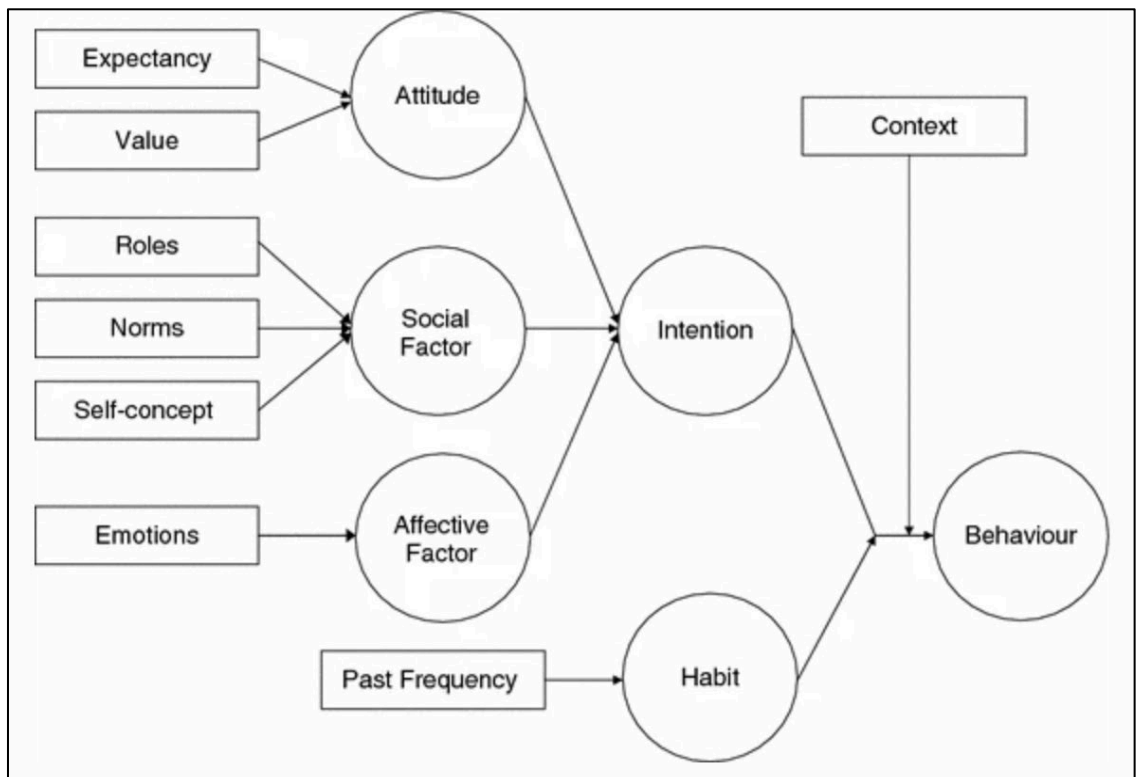

**Figure 1. Triandis' theory of interpersonal behaviour (Domarchi, et al., 2008)**

### 2.3. SURVEY STRUCTURE

The survey had four main sections – (i) DCE mode choice experiments; (ii) attitudes towards climate change and COVID-19; (iii) social factors affecting climate change and COVID-19 perceptions; and (iv) affective factors and other respondent characteristics of interest (e.g., demographics, commuting preferences and psychographics). Parts ii), iii) and iv) cover the main factors thought to affect travel behaviour, according to Triandis' theory. To gain comprehensive insights into respondents' views of climate change and COVID-19, survey questions covering all facets of Triandis' theory were devised.

## 3. SURVEY RESULTS

### 3.1. SAMPLE CHARACTERISTICS

**Figure 2** displays the gender and age representation achieved in the survey sample compared to national statistics for Scotland (National Records of Scotland, 2020). In total, 48.6% of the survey sample were male, 51.1% were female and 0.4% were non-binary. The most notable misrepresentations of age groups in the survey sample include an underrepresentation of those aged over 75, an overrepresentation of females aged 25–34 and 35–44, and an overrepresentation of males aged 65–74.

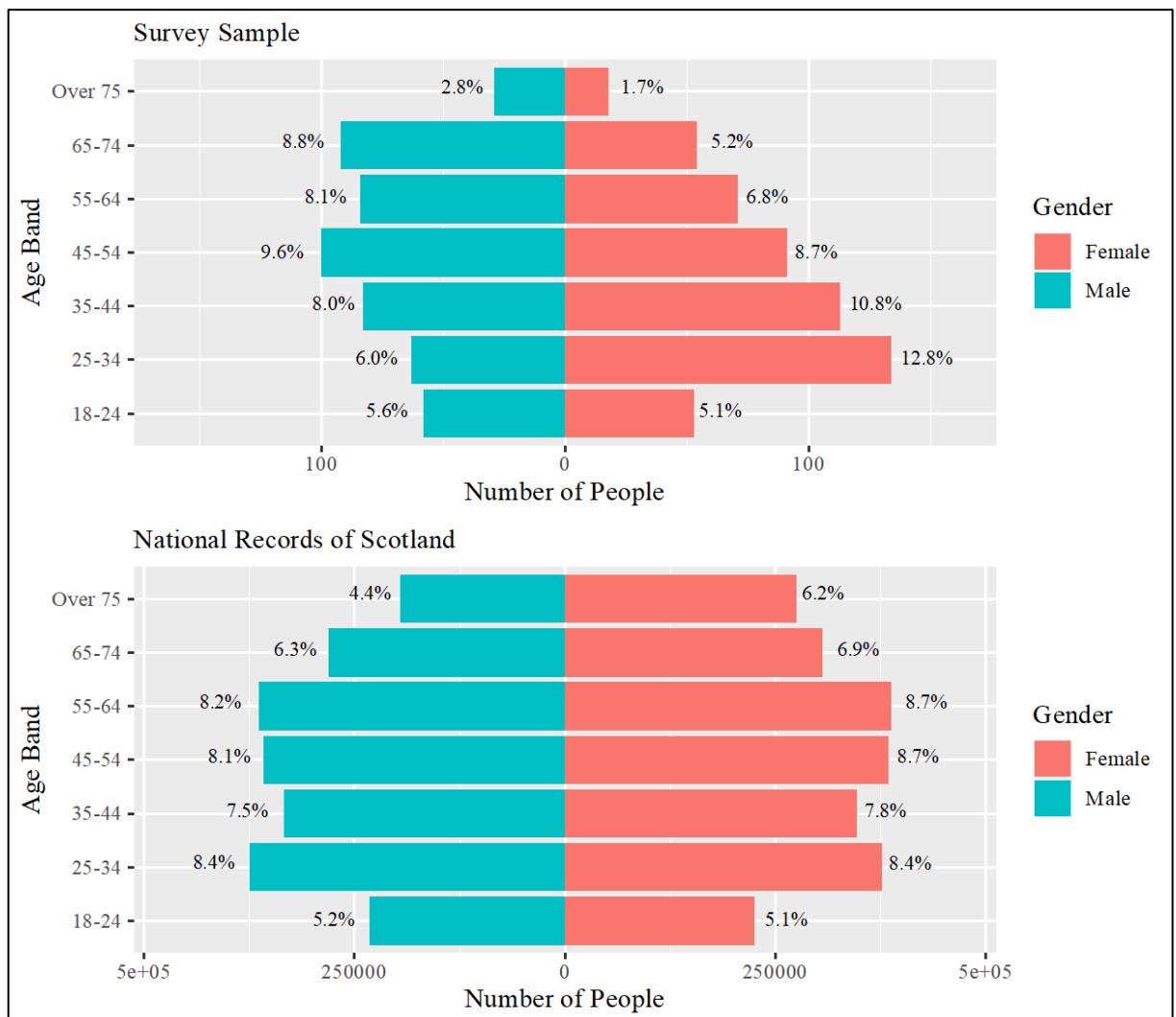

**Figure 2. Age-gender population pyramids for survey sample (n=1,046) versus National Records for Scotland (n=5,466,000)[1]**

**Figure 3** shows the distribution of household income (£ per annum before tax) among the survey respondents. Around 52% of respondents reported an annual income of either £10,001–20,000, £20,001–30,000 or £30,001–40,000, which is roughly consistent with the median household income in the UK – £29,900 per year (Office for National Statistics, 2020).

---

[1] It should be noted that four respondents identified as non-binary, three of whom were aged 18–24 and one who was aged 25–34.

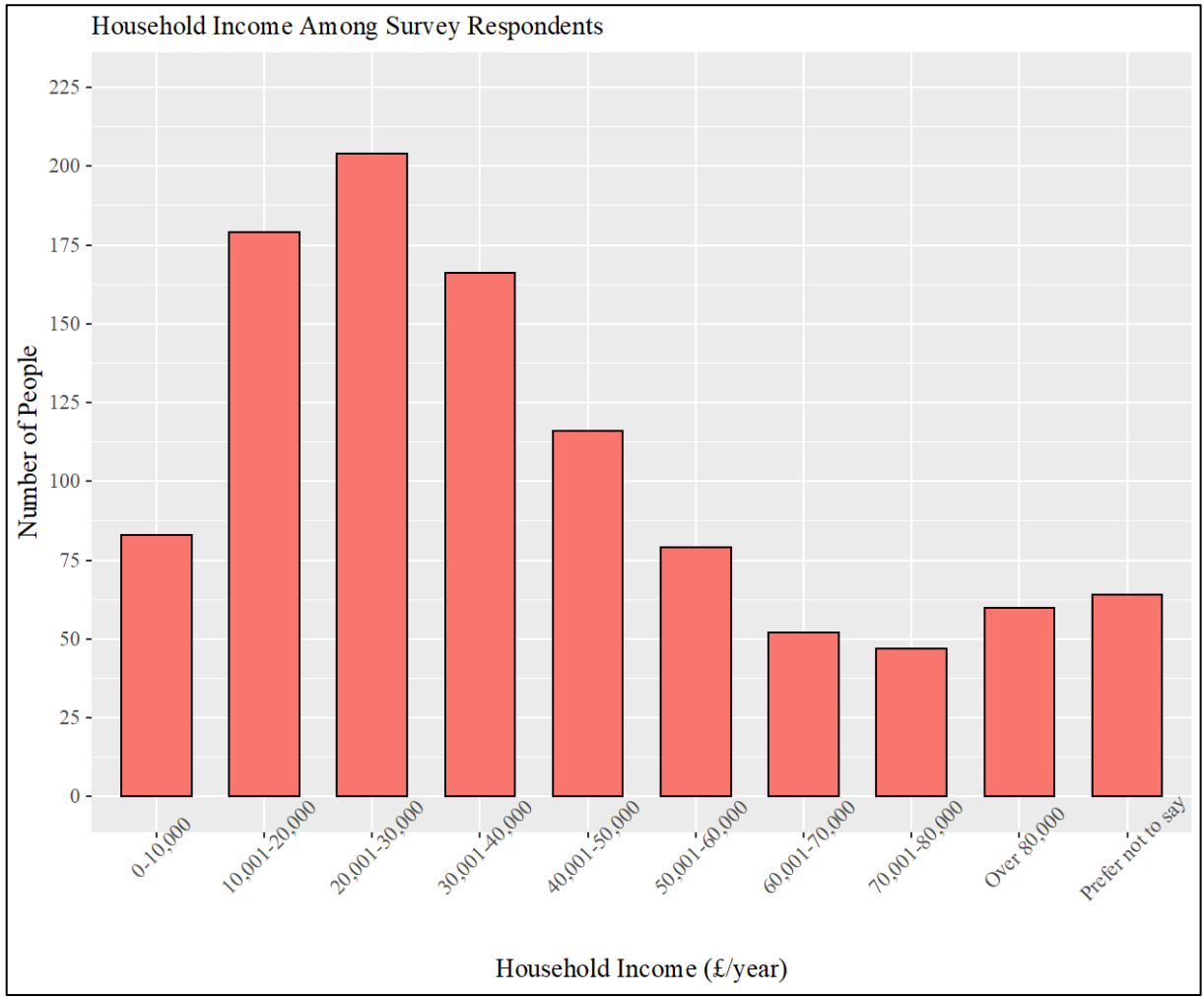

**Figure 3. Distribution of household income among survey respondents (n=1,050)**

### 3.2. CURRENT COMMUTING TRENDS

The survey gauged aspects of respondents' current commuting behaviour, therefore, most of the statistics in this section refer to the commuting contingent of the survey sample. Commuting information was provided by 502 respondents (47.8% of the total sample), whereas the remaining respondents reported that they complete all their work from home (19.7%) or are unemployed or not in part/full-time education (32.5%). **Figure 4** displays the time taken for the survey respondents to complete a one-way commuting trip, where the median value for the survey sample (25 minutes) is indicated by a red dashed line.

Transport Scotland (2020) data for typical one-way commuting trips is displayed in Figure 5. Transport Scotland record these data using intervals rather than a continuous scale, which was adopted in this study, therefore results are not directly comparable. According to Transport Scotland (2020), the majority of Scottish residents (~80%) have one-way commuting trip durations between 5 and 30 minutes. In comparison, approximately 65% of the survey sample reported one-way commuting trip durations between 5 and 30 minutes.

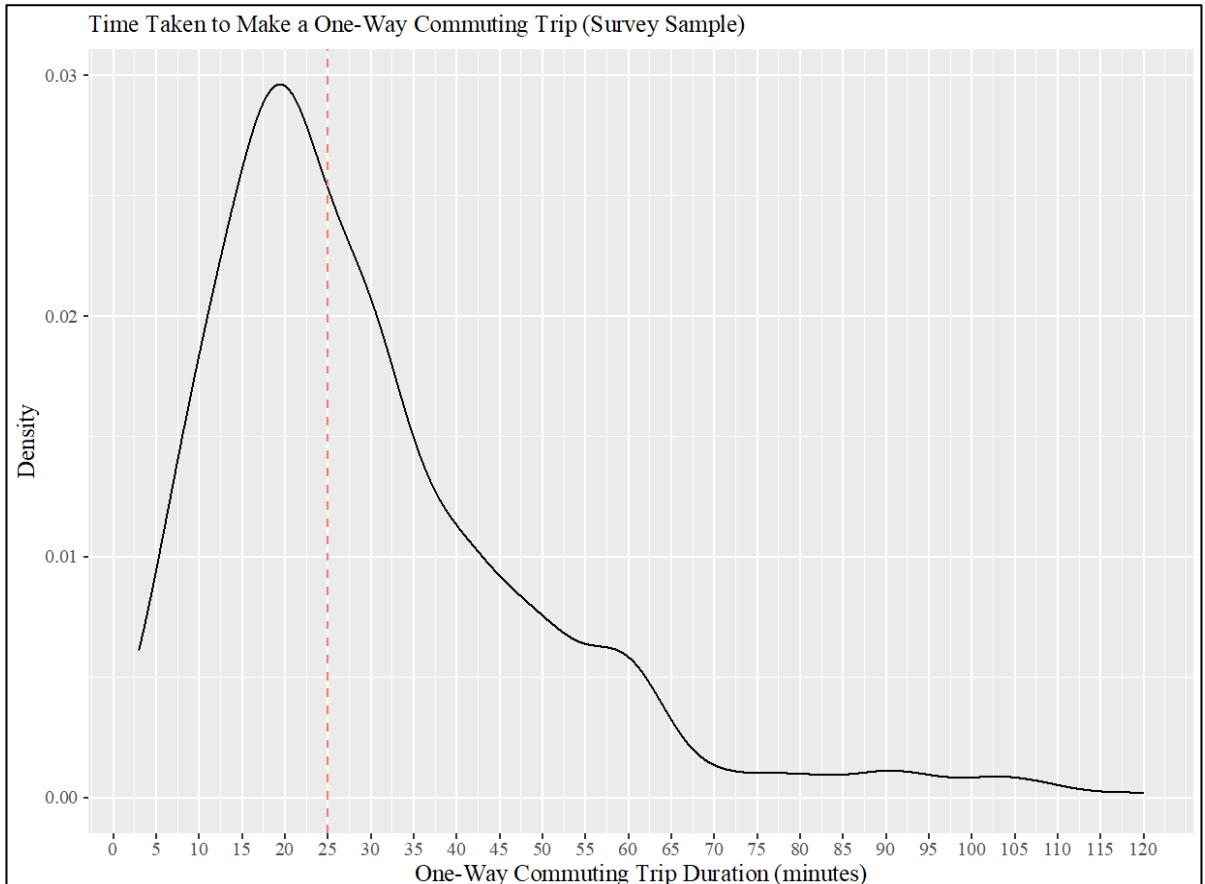

**Figure 4. One-way commuting trip duration among survey respondents, where the red dashed line indicates median trip duration (n=502)**

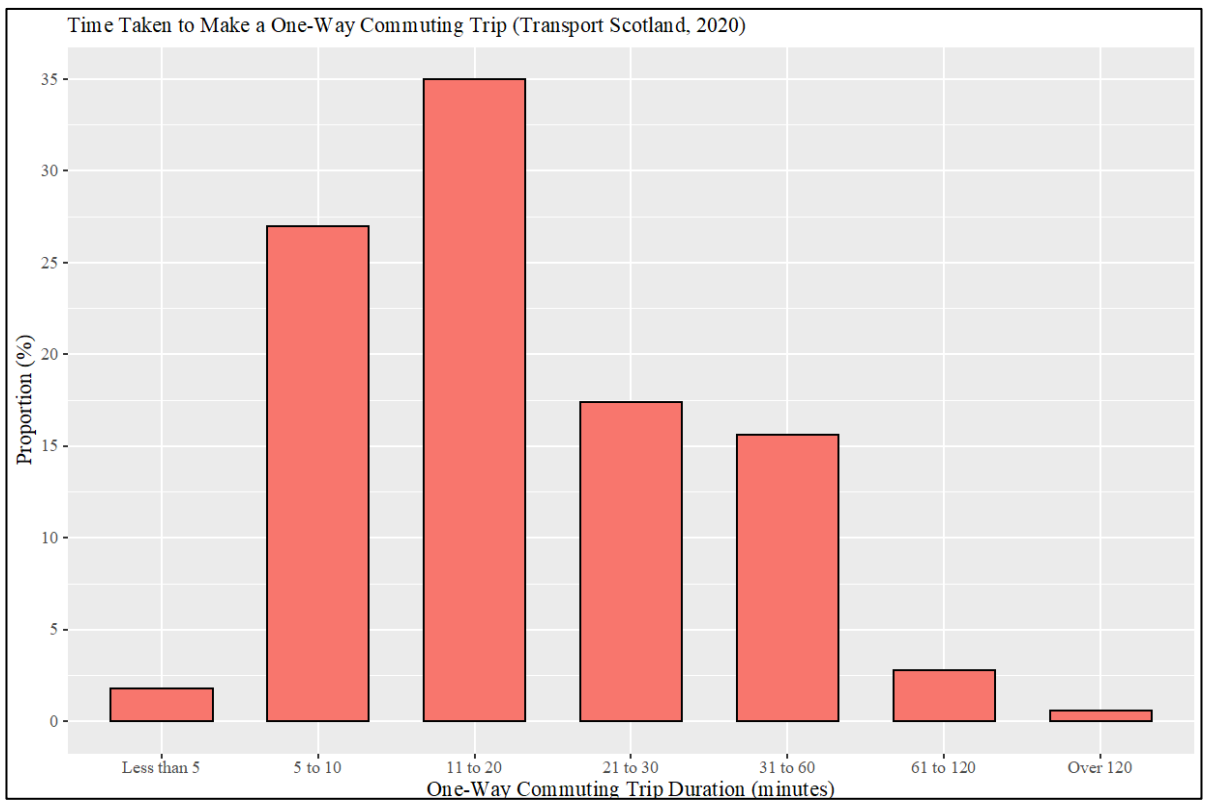

**Figure 5. One-way commuting trip duration among Scottish residents in 2020, as per Transport Scotland (2020)**

Survey respondents who commute were asked about the modes of transport they typically use for commuting trips (for full results see Appendix: **Table 1**).

- Around 30% of respondents commute by private vehicle (a car or van) five or more days per week, while ~69% of respondents make at least one commuting trip per week by private vehicle.
- A small proportion of respondents (~9%) make five or more commuting trips per week by bus and a smaller number (~3%) make five or more trips per week by train. ~27% of respondents make at least one weekly commuting trip by bus, compared to ~12% by train.
- Approximately 22% of respondents make five or more weekly commuting trips by active travel modes (~3% by bicycle and ~19% on-foot). Almost 70% of respondents commute actively at least one time per week (~18% by bicycle and ~52% on-foot).

**Figure 6** displays the frequency at which survey respondents completed certain activities that are likely to have been influenced by COVID-19, including working from home (WFH), online shopping for groceries and online shopping for non-grocery products (for full results see Appendix: **Table 2**).

- For the WFH question, 15.1% of respondents said that they currently WFH five or more days per week, 33.8% WFH at least one day per week, and 61.2% WFH less than once per month or never. This finding is comparable to pre-pandemic WFH statistics for Scotland. As expected, a lower proportion of the Scottish population (16.1%) worked from home in 2019 (Transport Scotland, 2019).
- A small minority of respondents shop online for groceries more than five or more days per week (1.3%), but a more considerable proportion (21.0%) of respondents shop online for groceries at least once per week. The majority of respondents (53.8%), however, shop online for groceries less than once a month or never.
- Interestingly, more respondents shop online for non-grocery products five days per week (3.1%) compared to online grocery shopping. 32.0% of respondents complete online non-grocery shops at least once per week, which is again greater than the proportion of those shopping online for grocery products. In this case, the majority of respondents (50.4%) said that they shop online for non-grocery products one to three times per month, while 17.6% said they do this less than once a month or never.

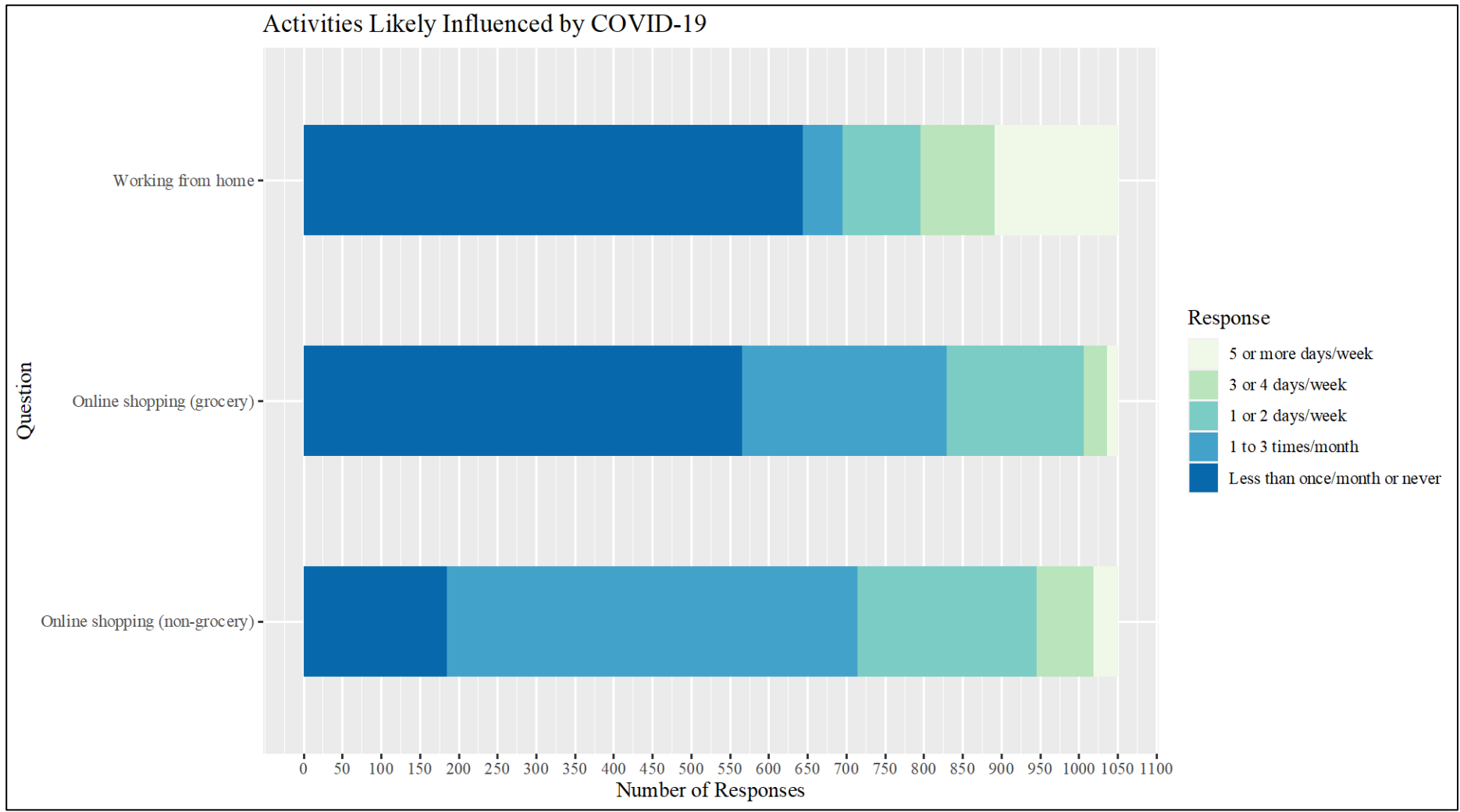

Figure 6. Distribution of responses to the survey question: "How often do you do each of the following activities?", i) Online shopping (groceries only), ii) Working from home and iii) Online shopping (non-grocery) (n=1,050)

### 3.3. CLIMATE CHANGE AND TRAVEL BEHAVIOUR

As part of the survey, respondents were asked a range of attitudinal and behavioural questions regarding climate change and COVID-19. Section 3.3. will discuss survey responses to climate change-related questions, whereas Section 3.4. displays the results for the equivalent COVID-19 survey questions. Appendix **Tables 3**, **4**, **5** and **7**, allow for direct comparisons between responses to equivalent questions for climate change and COVID-19.

#### 3.3.1. CLIMATE CHANGE: EXPECTATIONS, IMPORTANCE AND ASCRIPTION OF RESPONSIBILITY

This section will discuss survey responses to questions that gauged respondents' attitudes towards climate change, including expectations, importance and ascription of responsibility. Respondents were first asked about their expectations about whether society's "travel choices" or "general lifestyle choices" could have an impact on climate change. The verbatim question was as follows: "to what extent do you agree with the following statements regarding the possibility to have an impact on climate change?" (For full results see Appendix: **Table 3**). Key results include:

- 62.8% of respondents agreed or strongly agreed that society's travel choices can have an impact on climate change.
- A slightly smaller proportion of respondents (57.0%) agreed or strongly agreed that society's general lifestyle choices can have an impact on climate change.
- It is worth noting that some respondents remain to be convinced (i.e., answered neither agree nor disagree or disagreed to some extent) that travel choices (11.6%) and general lifestyle choices (13.7%) can have an impact on climate change.

Respondents were then asked a similar question regarding the importance of taking actions that reduce the effects of climate change. The distribution of responses to this survey questions is shown in **Figure 7** (for full results see Appendix: **Table 4**).

- 59.0% of respondents agreed or strongly agreed that it is important "to adopt travel habits that reduce climate change", while only 4.4% disagreed somewhat, disagreed or strongly disagreed. The remaining respondents either agreed somewhat (25.6%) or neither agreed nor disagreed (6.6%).
- A larger proportion of respondents (68.3%) agreed or strongly agreed that it is important "to take actions (in general) that reduce climate change", while 5.7% disagreed somewhat, disagreed or strongly disagreed. The remaining respondents either agreed somewhat (29.2%) or neither agreed nor disagreed (7.9%).

It may be that respondents who only agree somewhat, neither agree nor disagree, or who disagree to any extent, require more convincing that their travel habits and general actions are an important factor affecting climate change.

Respondents where then asked to ascribe responsibility for taking action to reduce climate change between themselves, others, industry and government. The results of this question are shown in **Figure 8** (for full results see Appendix: **Table 5**). The main findings were as follows:

- 56.0% of respondents agreed or strongly agreed with the statement "it is my responsibility to alter my lifestyle and reduce the negative effects of climate change".
- 26.5% of respondents agreed or strongly agreed with the statement "it is the Government's responsibility, not mine, to reduce the effects of climate change".
- 26.4% of respondents agreed or strongly agreed with the statement "it is the responsibility of industry, not mine, to reduce the effects of climate change".
- 13.4% of respondents agreed or strongly agreed with the statement "it is other people's responsibility, not mine, to reduce the effects of climate change".

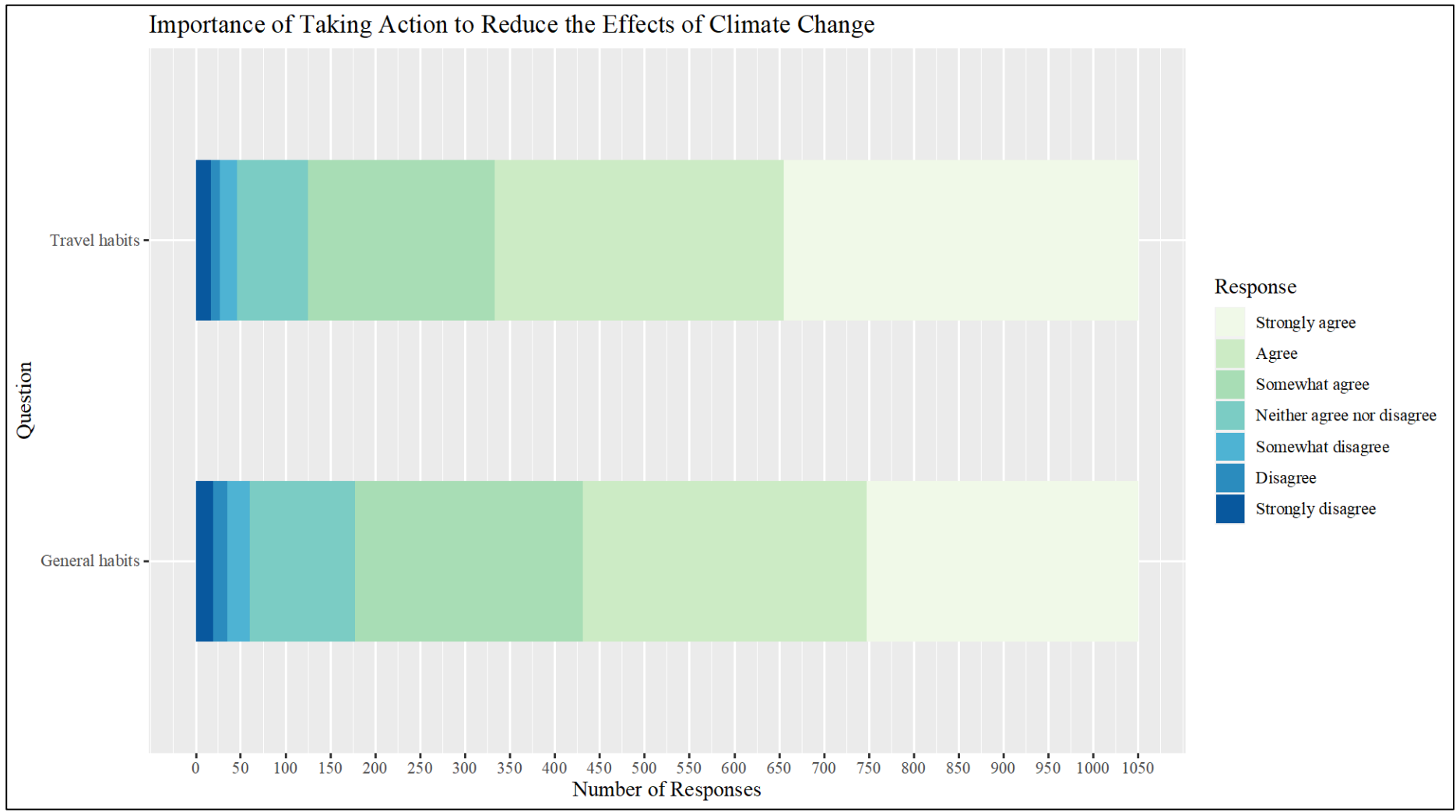

Figure 7. Distribution of responses to the survey question: "To what extent do you agree with the following statements regarding the importance of doing something to reduce the effects of climate change?", i) "It is important to adopt travel habits that reduce climate change" and ii) "It is important to take actions (in general) that reduce climate change" (n=1,050)

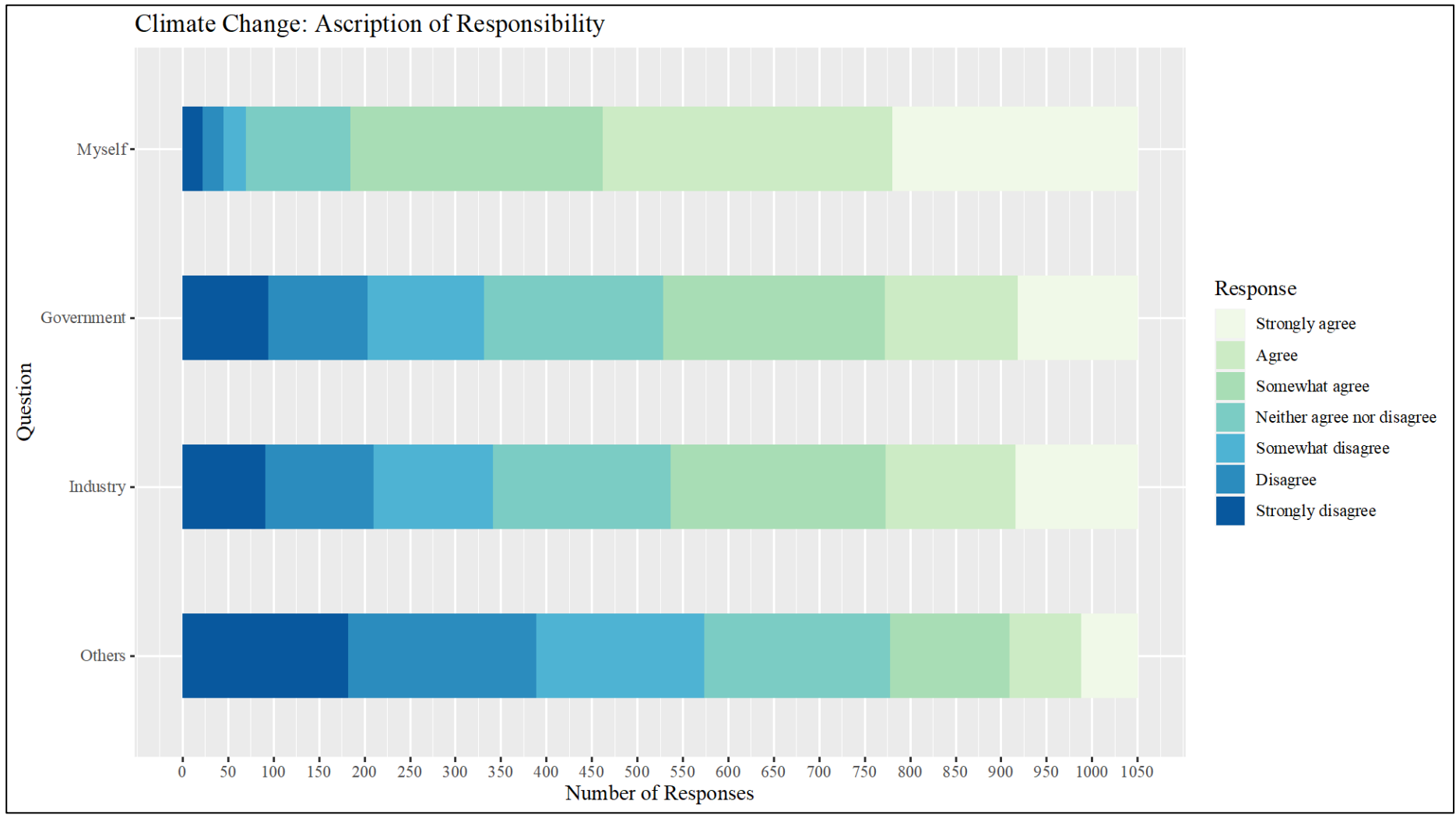

Figure 8. Distribution of responses to the survey question: "To what extent do you agree with the following statements about the responsibility of acting to reduce climate change?" (n=1,050

Interestingly, the survey respondents felt that they themselves are most responsible for altering their own lifestyle to reduce the negative effects of climate change. However, this finding should be interpreted with caution given that the statements for government, industry and other people included the phrase "government's/industry's/other people's responsibility, not mine", therefore respondents may have not agreed with this statement on the basis that it suggests they are not personally responsible at all, e.g., someone may consider themselves and government equally responsible.

### 3.3.2. CLIMATE CHANGE: OPENNESS TO CHANGE AND SOCIAL FACTORS

This section explores respondents' openness to change, i.e., their willingness to adopt pro-environmental behaviours, as well as exploring climate-related social factors. As shown in **Figure 9**, respondents were asked how likely it was that they would adopt various pro-environmental behaviours (for full results see Appendix: **Table 6**). The behaviours in **Figure 9** are listed in descending order from most to least likely (in terms of mean response). The most popular outcome was recycling more, followed by conserving energy in the home, more active travel (i.e., on-foot or by bicycle) and making less flights. It is likely that these actions receive widespread support because they are typically free of cost, or may even result in reduced costs, and are also easier to adopt than some of the less popular outcomes.

In descending order, the next most popular response was switching to an electric vehicle, followed by using more public transport, eating less meat, buying organic products and joining an environmental group. It is likely that switching to an electric vehicle and buying organic products received low support because both of these actions are typically more expensive than alternative products in their respective markets. Eating less meat and joining an environmental group are both politically charged issues that often divide opinion, which likely explains the relatively low levels of support for these outcomes. It should be noted that joining an environmental group was the only outcome where a majority of respondents (64.0%) considered their participation to be extremely unlikely, unlikely, or somewhat unlikely. In all remaining questions, the proportion of respondents who responded unlikely (to any extent) ranged from 2.8%–45.9%.

The next survey questions explored social factors (social norms, social responsibility and perceived social capability) surrounding the issue of climate change. Respondents were first asked about social norms, and specifically what people important to them do and think about climate change (for full results see Appendix: **Table 7**). Key results were as follows:

- 22.6% of respondents agreed or strongly agreed with the statement "most people who are important to me, are personally doing something to help reduce the risk of climate change". The bulk of respondents, however, agreed somewhat (30.6%) or neither agreed nor disagreed (26.9%), while the remaining 20.0% disagreed to some extent.
- In contrast, more respondents (37.1%) agreed or strongly agreed with the statement "it is generally expected of me that I do my bit to help reduce the risk of climate change", while 34.5% agreed somewhat, 17.2% neither agreed nor disagreed and 11.2% disagreed to some extent.

Respondents were then asked about social responsibility, in other words, they were asked what they felt their role was in addressing climate change (for full results see Appendix: **Table 7**). Results were as follows:

- 27.3% of respondents agreed or strongly agreed with the statement "I feel that I should inspire people to take action to reduce the effects of climate change", 20.6% agreed somewhat, 29.3% neither agreed nor disagreed, and 22.8% disagreed to some extent.
- A lower proportion (11.3%) agreed or strongly agreed with the statement "I am often asked for advice by other people about ways to reduce the effects of climate change" and the majority of respondents (57.4%) disagreed to some extent.

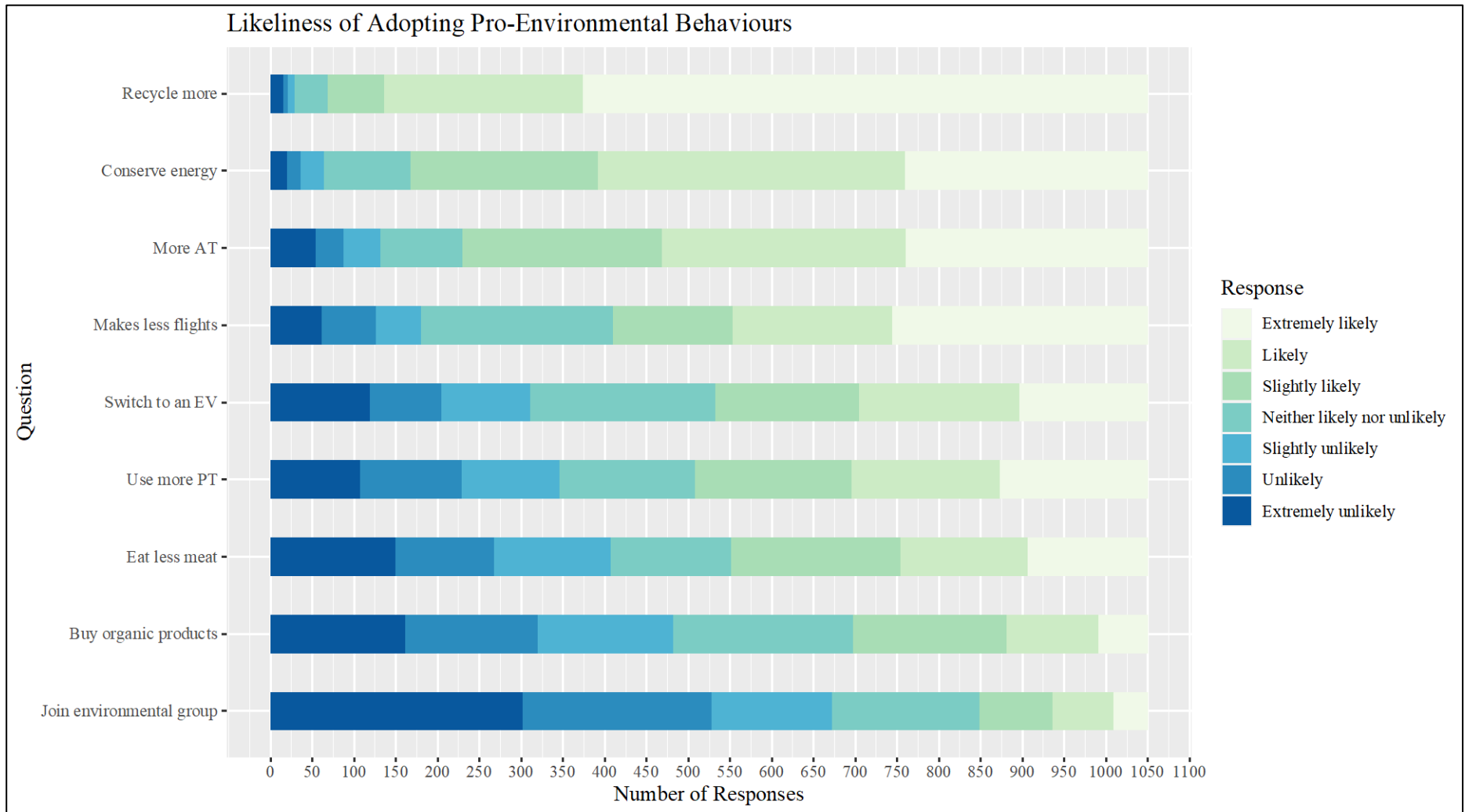

Figure 9. Distribution of responses to the survey question: "How likely is it that you would adopt any of the following actions to reduce your carbon footprint?" (n=1,050)

Respondents were then asked about their perceived social capability, regarding whether they felt they could influence other people's actions in a way that reduces the effects of climate change (for full results see Appendix: **Table 7**). Key findings were as follows:

- 13.81% of respondents agreed or strongly agreed with the statement "I feel that I am able to inspire people to take action and reduce the effects of climate change", while 15.6% agreed somewhat and 29.2% neither agreed nor disagreed. The remaining 41.4% of respondents disagreed to some extent.

## 3.4. COVID-19 AND TRAVEL BEHAVIOUR

### 3.4.1. COVID-19: EXPECTATIONS, IMPORTANCE AND ASCRIPTION OF RESPONSIBILITY

This section displays respondents' answers to COVID-19-related survey questions. As was the case for climate change survey questions, respondents were asked about expectations (i.e., effectiveness of actions), importance of taking action and ascription of responsibility, in the context of reducing the spread of infectious diseases, such as COVID-19. Results for questions gauging COVID-19 expectations are summarised below (for full results see Appendix: **Table 3**):

- 59.8% of respondents agreed or strongly agreed that society's travel choices can have an impact on the spread of infectious diseases, 26.1% agreed somewhat, 7.7% neither agreed nor disagreed and 6.4% disagreed to some extent.
- A larger proportion of respondents (65.4%) agreed or strongly agreed that society's general lifestyle choices can have an impact on the spread of infectious diseases. The remaining respondents agreed somewhat (23.3%), neither agreed nor disagreed (6.7%) or disagreed to some extent (4.6%).
- Interestingly the trend for the equivalent climate change questions was reversed for COVID-19 (see section 3.3.1. and Appendix: **Table 3**), in other words, respondents believe society's general lifestyle choices have greater impact on the spread of infectious diseases than travel habits.

Respondents were then asked about the importance of taking action to reduce the spread of infectious diseases like COVID-19 (see Appendix: **Table 4** for full results and comparison to climate change importance questions). Key findings are displayed in **Figure 10** and discussed below:

- 73.5% of respondents agreed or strongly agreed with the statement "it is important to take actions that reduce the spread of infectious diseases like COVID-19", 17.8% agreed somewhat, 5.4% neither agreed nor disagreed and 3.2% disagreed to some extent.
- A smaller proportion of respondents (68.2%) agreed or strongly agreed with the statement "it is important to adopt travel habits that reduce the spread of infectious diseases like COVID-19", 19.0% agreed somewhat, 8.2% neither agreed nor disagreed and 4.7% disagreed to some extent.
- As was the case for the question gauging COVID-19 expectations, respondents tend to think that society's general lifestyle choices are a more important factor than travel habits. A likely explanation is that respondents recognise that both travel habits and general lifestyle choices are important factors affecting COVID-19 infection rates, however, lifestyle choices are marginally more important as they account for a wider variety of behaviours (e.g., social contact, mask wearing etc.).

Respondents were then asked about their views, regarding ascription of responsibility when trying to limit the spread of infectious diseases like COVID-19. **Figure 11** shows the distribution of responses for this question. The main findings were as follows (see Appendix: **Table 5** for full results, and comparison to equivalent ascription of responsibility question for climate change):

- 62.2% of respondents agreed or strongly agreed with the statement "it is my responsibility to alter my lifestyle to reduce the spread of infectious diseases like COVID-19", while 23.6% agreed somewhat, 8.8% neither agreed nor disagreed, and the remaining 5.4% disagreed to some extent.
- 23.7% of respondents agreed or strongly agreed with the statement "it is the Government's responsibility, not mine, to reduce the spread of infectious diseases like COVID-19". A considerable proportion of respondents (36.7%) disagreed to some extent while the remainder agreed somewhat (20.1%) or neither agreed nor disagreed (19.5%).
- 19.1% of respondents agreed or strongly agreed with the statement "it is other people's responsibility, not mine, to reduce the spread of infectious diseases like COVID-19". A considerable proportion of respondents (46.3%) disagreed to some extent again, 16.4% agreed somewhat and 18.3% neither agreed nor disagreed.

Respondents tended to hold themselves most responsible for altering their lifestyle to reduce the spread of infectious diseases like COVID-19, however, this trend may be partially induced by the wording of the question statements, as discussed for the equivalent climate change questions in Section 3.3.1. Despite this, some interesting comparisons can be made between the ascription of responsibility questions for climate change and COVID-19 (see Appendix: **Table 5**).

85.8% of respondents agree to some extent that they are responsible for reducing the spread of infectious diseases, which is slightly more than those who agree to some extent that they are responsible for altering their lifestyle and reducing the effects of climate change (82.5%). Respondents also ascribed greater responsibility to the Government in the context of climate change – 49.7% agreed to some extent that the Government were responsible for reducing the negative effects of climate change compared to 43.8% when asked about reducing the spread of infectious diseases. This suggests that the respondents tend to take more personal responsibility for reducing the spread of infectious diseases. This may be because there have been myriad public health information campaigns informing the public of ways to reduce the spread of COVID-19, whereas there may be relatively lower knowledge of the ways in which individuals can reduce their impact on climate change.

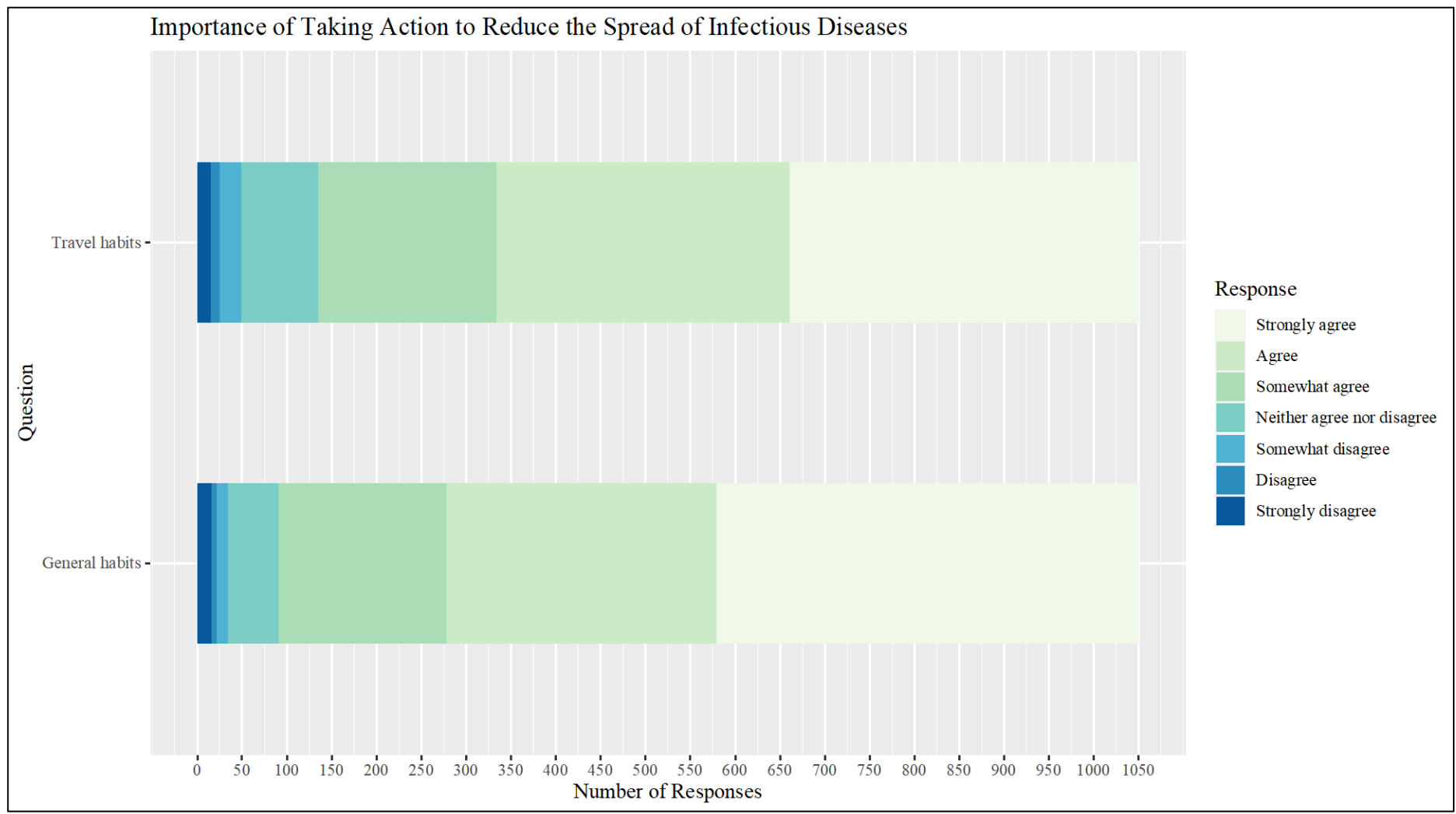

Figure 10. Distribution of responses to the survey question: "To what extent do you agree with the following statements regarding the importance of doing something to reduce the spread of infectious diseases like COVID-19?" i) "It is important to adopt travel habits that reduce the spread of infectious diseases like COVID-19" and ii) "It is important to take actions that reduce the spread of infectious diseases like COVID-19" (n=1,050)

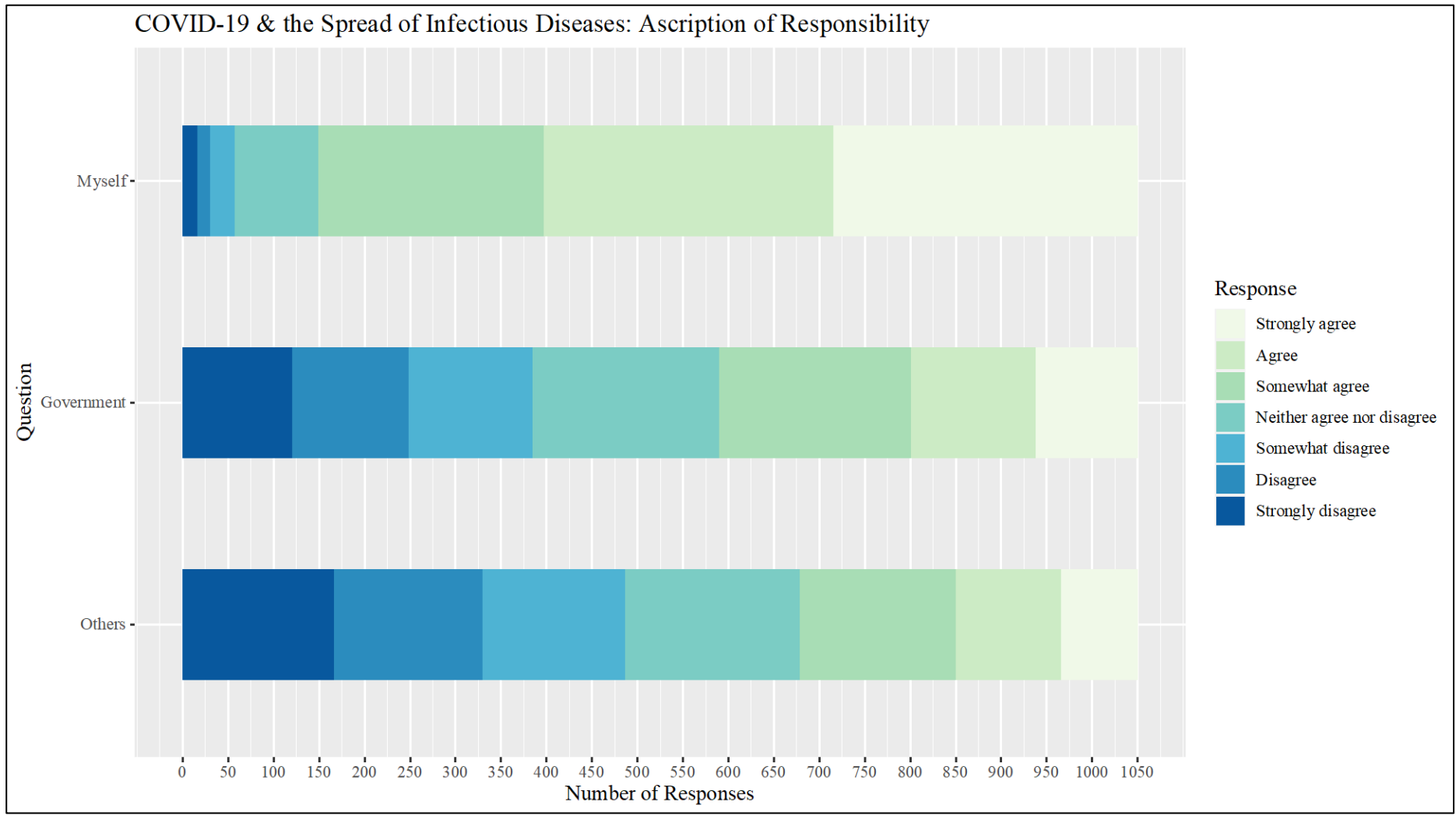

Figure 11. Distribution of responses to the survey question: "To what extent do you agree with the following statements about the responsibility of acting to reduce impact of infectious diseases like COVID-19?" (n=1,050)

### 3.4.2. COVID-19: OPENNESS TO CHANGE AND SOCIAL FACTORS

This section explores respondents' openness to adopt behaviours that reduce the spread of infectious diseases, as well as the social factors affecting COVID-19 perceptions. **Figure 12** shows responses to a survey question gauging how likely it was that respondents would engage with different behaviours that reduce the spread of infectious diseases when COVID-19 is in general circulation (i.e., level three of the UK Government's alert system). All outcomes in **Figure 12**, excluding volunteering in the NHS, received majority support (i.e., answered slightly likely, likely, or extremely likely) from the respondents. The level of support for all outcomes other than volunteering in the NHS ranged from 58.5% to 86.5% (for full results see Appendix: **Table 8**).

The highest level of support was received for following official government guidelines, followed by more frequent WFH, more active travel and more online shopping. The next most popular outcomes were making less flights, followed by reducing travel, reducing social contact, using less public transport, and volunteering in the NHS. As expected, respondents overwhelmingly support simple measures to reduce the spread of infectious diseases, such as following government guidelines and WFH more. In terms of travel-related behaviours, respondents were most likely to select more active travel, followed by making less flights, reducing travel in general and using less public transport.

The next survey questions gauged the social factors affecting COVID-19 behaviours. These questions are comparable with the equivalent social factors questions for climate change, discussed in Section 3.3.2. Respondents were first asked about social norms in the context of COVID-19 (for full results and comparison with climate change questions see Appendix: **Table 7**). Key results were as follows:

- 47.3% of respondents agreed or strongly agreed with the statement "most people who are important to me, are personally doing something to reduce their risk of contracting/spreading infectious diseases like COVID-19", while a considerable proportion agreed somewhat (32.7%). The remaining respondents neither agreed nor disagreed (12.2%) or disagreed to some extent (7.8%).
- 56.0% of respondents agreed or strongly agreed with the statement "it is generally expected of me that I do my bit to reduce the risk of contracting/spreading infectious diseases like COVID-19", while a large number of respondents agreed somewhat (30.4%), 9.0% neither agreed nor disagreed and 4.7% disagreed to some extent.
- Although the majority of respondents agree to some extent with both statements, it is perhaps surprising that around 14% of respondents (who answered neither agree nor disagree or disagreed to some extent) do not feel that they are expected to act in a way that reduces the spread of COVID-19.

Respondents were then asked about social responsibility in the context of reducing the spread of infectious diseases like COVID-19 (for full results see Appendix: **Table 7**). Results were as follows:

- 28.1% of respondents agreed or strongly agreed with the statement "I feel that I should inspire people to take action and reduce their risk of contracting/spreading infectious diseases like COVID-19", 22.9% agreed somewhat, 27.0% neither agreed nor disagreed and 22.1% disagreed to some extent.
- 15.9% of respondents agreed or strongly agreed with the statement "I am often asked for advice by other people about ways to reduce their risk of contracting/spreading infectious diseases like COVID-19", 12.6% agreed somewhat and 24.7% neither agreed nor disagreed. In this case, a large proportion of respondents disagreed to some extent (46.9%) with 14.1% of these respondents in strong disagreement.
- Although a slight majority of respondents agree to some extent that they should inspire people to take action to reduce the spread of COVID-19, in reality only a minority of respondents are asked for advice.

Respondents were next asked a question about their perceived social capability when trying to inspire others to take action to reduce the spread of COVID-19. Key results were as follows (for full results see Appendix: **Table 7**):

- 17.2% of respondents agreed or strongly agreed with the statement "I feel that I am able to inspire people to take action and reduce their risk of contracting/spreading infectious diseases like COVID-19", 21.2% agreed to some extent, 29.0% neither agreed nor disagreed and 32.6% disagreed to some extent.

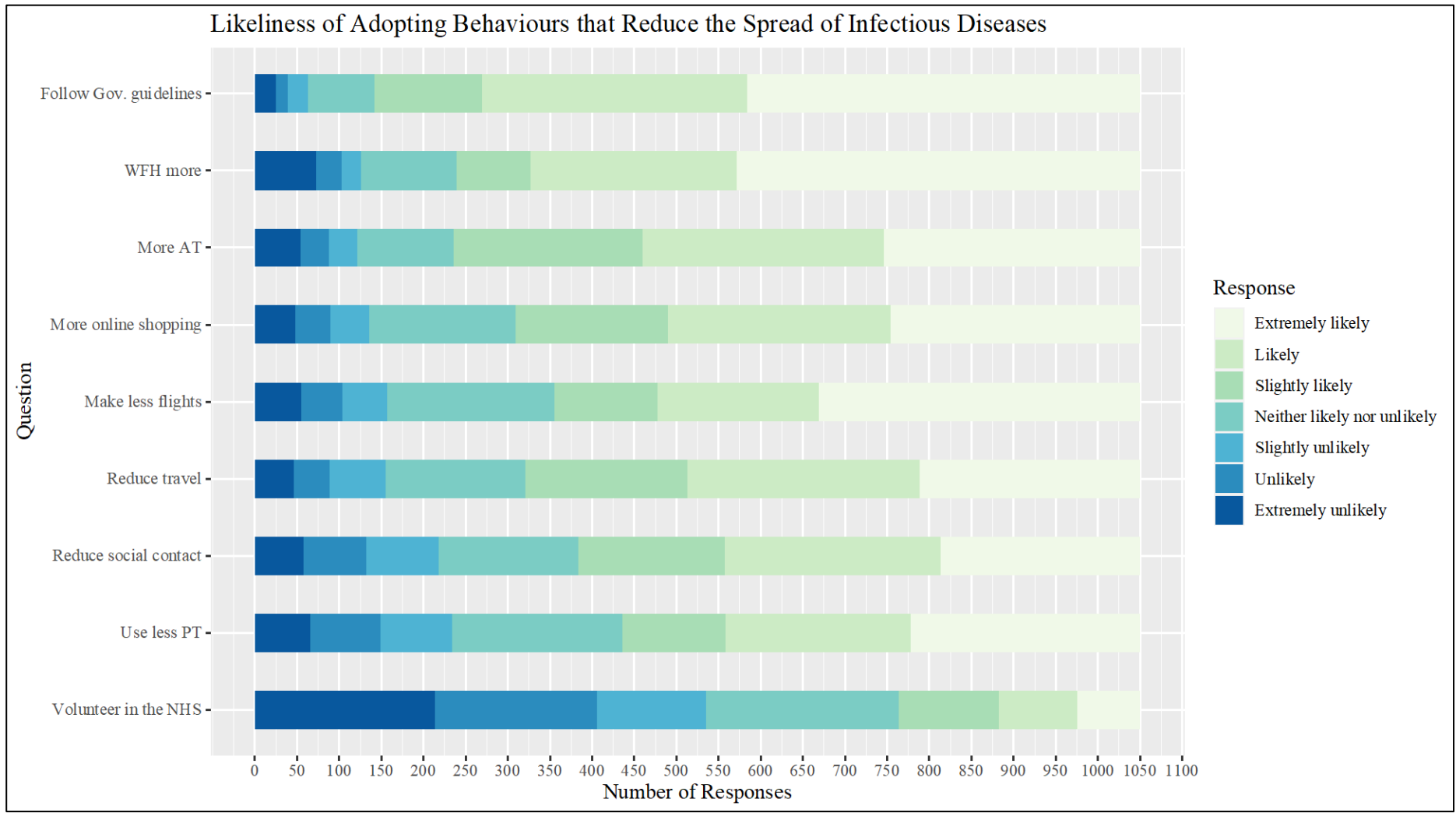

Figure 12. Distribution of responses to the survey question: "How likely is it that you would adopt any of the following actions to reduce the impacts of pandemics like COVID-19 when the virus is in general circulation (level three of the UK Government's alert system)" (n=1,050)

## 3.5. MODE PREFERENCES FOR DIFFERENT COVID-19 & WFH SCENARIOS

As discussed previously, the survey included a discrete choice experiment to explore respondents' mode preferences for commuting trips in four different scenarios. The scenarios were defined as follows: (I) low WFH (no more than four days per month) and no COVID-19 risk (Level 1 of the UK Government's COVID-19 alert system); (II) low WFH (no more than four days per month) and medium COVID-19 risk (Level 3 of the UK Government's COVID-19 alert system); (III) high WFH (three or more days per week) and no COVID-19 risk; and (IV) high WFH (three or more days per week) and medium COVID-19 risk. Respondents were asked to select their preferred mode of travel – either car, bus or bicycle – for commuting trips in each scenario. The choices of respondents were also dependent on travel time, travel cost and carbon emissions associated with each mode, however, only differences between mode preferences across the scenarios will be discussed in this report.

**Figure 13** shows respondents' mode choices per scenario[2]. Scenario 1 may be viewed as a pre-COVID-19 baseline, given that respondents were asked to consider an environment where COVID-19 poses no risk and WFH is only possible for a maximum of four days per month. Scenario 2 considered a setting where WFH is infrequent, but COVID-19 poses a medium risk. As may be expected, respondents were less likely to favour travel by bus in Scenario 2 (down ~6% percentage points compared to Scenario 1) and more likely to select private modes of transport, i.e., car or bicycle. Intuitively this makes sense, as some respondents would likely prefer to minimise their infection risk through avoiding crowded and enclosed spaces.

In Scenario 3, where WFH was expected three of more days per week and COVID-19 poses no threat, 32.9% of respondents chose bicycle as their preferred mode, which remained roughly consistent in Scenario 4, where there was high WFH and medium COVID-19 risk. However, bicycle was a considerably more popular choice in the high WFH scenarios (Scenarios 3 and 4), in comparison to the low WFH scenarios (Scenarios 1 and 2). Only 47% of respondents chose car in Scenario 3, which is lower than any other scenario. Through comparison with Scenario 1, this suggests that more frequent WFH, when no risk is posed by COVID-19, may increase the appeal of commuting by bicycle and reduce the appeal of travelling by car. 20.2% of respondents chose bus in Scenario 3, which is marginally less than in Scenario 1. This suggests that high WFH could modestly reduce commuting trips by bus.

In Scenario 4, where WFH was expected three of more days per week and COVID-19 poses a medium risk, 32.7% of respondents chose bicycle. This suggests that WFH is the main factor affecting bicycle choices, whereas the effect of COVID-19 risk level is negligible. The main differences in Scenario 4, in comparison to Scenario 3, were a rise in the popularity of cars and a decrease in the popularity of buses. Similar to the differences observed between Scenarios 1 and 2, respondents seem more hesitant to travel by bus when COVID-19 poses a medium risk and instead opt for private modes (car or bicycle).

In summary, increased COVID-19 infection risk appears to coax respondents towards private modes of travel and away from public transport. In the low WFH scenarios greater COVID-19 risk increases the shares of both cars and bicycles; in the high WFH scenarios greater COVID-19 risk has a negligible effect on the share of bicycles, and the shift is mainly from buses to cars. This shift from public transport to private vehicles has been evidenced by various studies throughout the pandemic (Jenelius & Cebecauer, 2020; Przybylowski, et al., 2021). Both scenarios where WFH was frequent (i.e., Scenarios 3 and 4), saw rises in the popularity of bicycles, when compared to the infrequent WFH scenarios (i.e., Scenarios 1 and 2). This suggests that Scottish residents may be more willing to commute by bicycle if they have to travel to their workplace only one or two days per week. The findings also suggest that frequent WFH may lead to modest decreases of commuting trips made by car or bus.

---

[2] It should be noted that there are 4,200 responses per scenario because respondents made four choices in each scenario (i.e., 1,050*4).

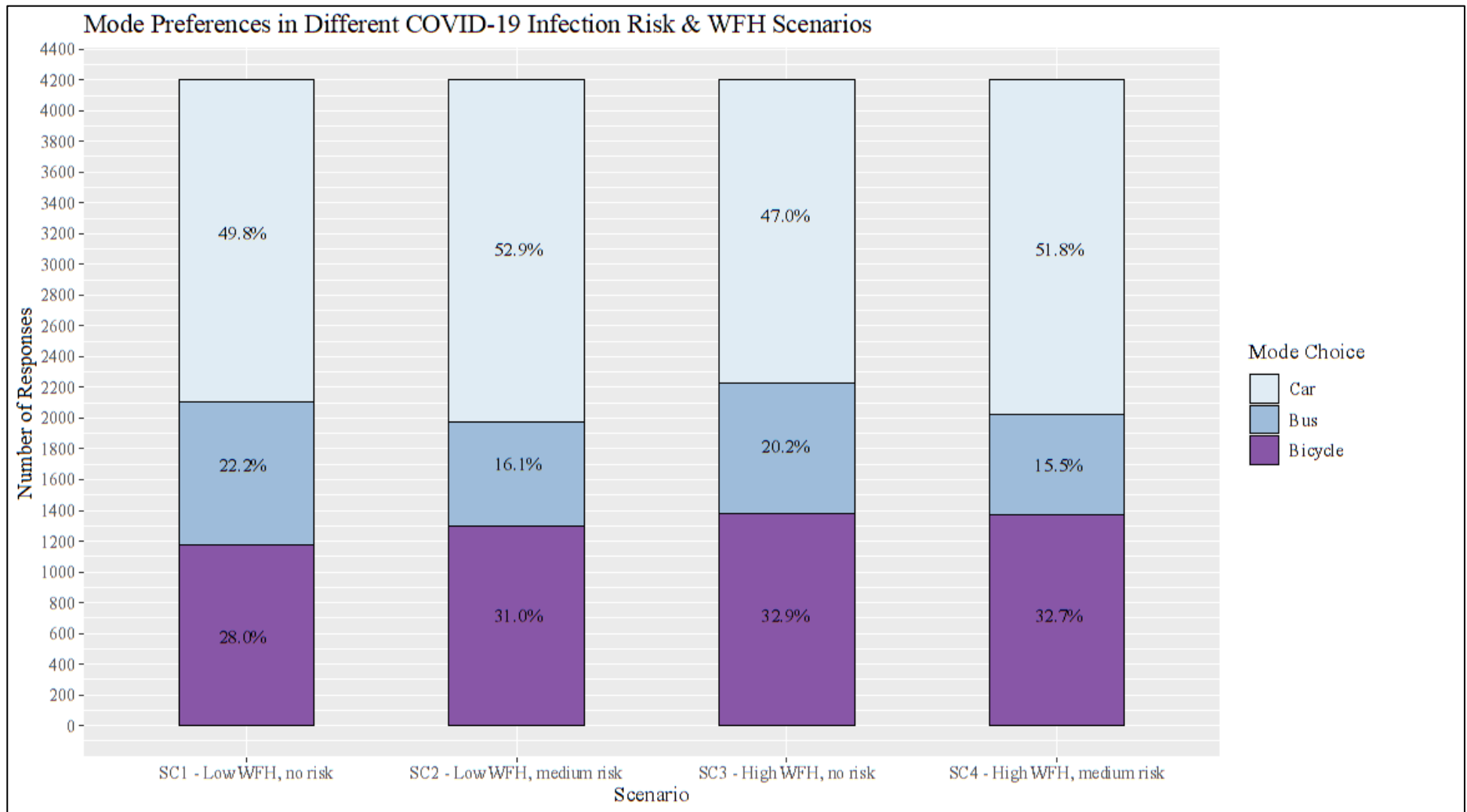

Figure 13. Modal preferences per COVID-19 infection risk and WFH scenario, where Scenario 1 = WFH no more than 4 days/month and no COVID-19 infection risk (Level 1 of the UK Government's COVID-19 alert system); Scenario 2 = WFH no more than 4 days/month and medium COVID-19 risk (Level 3 of the UK Government's COVID-19 alert system); Scenario 3 = WFH 3 or more days/week and no COVID-19 infection risk; and Scenario 4 = WFH 3 or more days/week and medium COVID-19 risk

## 4. CONCLUSIONS

To summarise, 17.5% of respondents must still be convinced (i.e., responded "neither agree nor disagree" or disagreed to some extent) that it is their responsibility to contribute to reducing the effects of climate change. A smaller proportion (11.6%) must be convinced about the effectiveness of changing travel habits to reduce the effects of climate change, while slightly more respondents (16.9%) still need to be convinced that altering their travel habits is an important factor affecting climate change. In terms of willingness to adopt pro-environmental travel behaviours, respondents were highly likely to use active travel modes more frequently and reduce their number of flights. Respondents were less likely to switch to an electric vehicle or to use more public transport, however, in both cases around half the respondents still said that they would be willing to adopt either. All of the aforementioned points may be addressed through suitable policy interventions, for example, public information campaigns to ensure the public understand the effectiveness and importance of changing travel habits to reduce the effects of climate change or grants to improve the affordability of electric vehicles.

14.2% of respondents were not convinced that they have any personal responsibility to reduce the spread of COVID-19, hence, the remaining 85.8% do take personal responsibility. The tendency of respondents to assume personal responsibility for reducing the spread of COVID-19 was also the case for reducing the effects of climate change, which possibly points to existence of people who lack trust in official science. The effectiveness and importance of travel choices in the context of reducing the spread of COVID-19 are comparable to the equivalent questions for climate change. For COVID-19, 14.1% of respondents still need to be convinced that travel choices can have an effect on the spread of COVID-19, whereas 12.9% of respondents remain to be convinced that travel choices are an important factor affecting the spread of COVID-19. Respondents were willing to support most behaviours that reduce the spread of infectious diseases or COVID-19, with the only exception being volunteering in the NHS. In terms of travel-related behaviours that reduce the spread of COVID-19, respondents were more likely to reduce their overall travel than using public transport less. More active travel was also more popular among respondents than reducing public transport use.

The DCE mode choice scenarios showed that the use of cars for commuting is relatively inelastic depending on frequency of WFH, however, bicycles become more popular in high WFH scenarios compared to low WFH scenarios. As expected, the attractiveness of public transport decreases with increased COVID-19 risk and private modes become more popular. The shift from bus to car is particularly pronounced when in a high WFH and medium COVID-19 risk scenario, in comparison to the high WFH and no COVID-19 risk scenario. These findings may be used to aid transport policy through subtly different phases of future pandemics.

**Table 1. Mode of travel preferences for commuting trips (n=502, i.e., the portion of the survey sample that regularly commutes)**

| Question | 5 or more days a week | 3 or 4 days a week | 1 or 2 days a week | 1 to 3 times a month | Less than once a month and more than once a year | Less than once a year or never |
|---|---|---|---|---|---|---|
| Private car or van (as driver or as a passenger) | 29.88% | 22.71% | 16.33% | 5.98% | 7.17% | 17.93% |
| Taxi, Uber or similar services, private hire vehicle | 1.00% | 1.39% | 4.78% | 11.55% | 18.92% | 62.35% |
| Bus, minibus or coach | 8.57% | 8.37% | 9.76% | 9.76% | 15.74% | 47.81% |
| Train | 2.59% | 3.59% | 5.78% | 9.96% | 22.11% | 55.98% |
| Bicycle | 3.19% | 4.78% | 10.16% | 6.77% | 10.76% | 64.34% |
| Walking (more than a quarter of a mile) | 18.73% | 17.13% | 15.74% | 11.35% | 6.57% | 30.48% |
| Subway | 0.80% | 1.20% | 3.39% | 4.98% | 7.57% | 82.07% |

**Table 2. Working from home, online shopping for groceries and online shopping for non-groceries (n=1,050)**

| Question | 5 or more days a week | 3 or 4 days a week | 1 or 2 days a week | 1 to 3 times a month | Less than once a month or never |
|---|---|---|---|---|---|
| Telecommute (work from home) | 15.14% | 9.14% | 9.52% | 4.95% | 61.24% |
| Shopping online to purchase non-grocery products (e.g., Amazon) | 3.05% | 6.95% | 22.00% | 50.38% | 17.62% |
| Shopping online to purchase grocery products (e.g., home delivery services) | 1.33% | 2.86% | 16.86% | 25.14% | 53.81% |

Table 3. Climate change & COVID-19 attitudes (expectations, n=1,050) (Climate change = green fill; COVID-19 = blue fill)

| Question | Strongly disagree | Disagree | Somewhat disagree | Neither agree nor disagree | Somewhat agree | Agree | Strongly agree |
|---|---|---|---|---|---|---|---|
| Society's general lifestyle choices can contribute to reducing climate change | 1.81% | 1.62% | 2.38% | 7.90% | 29.24% | 29.43% | 27.62% |
| Society's general lifestyle choices can contribute to reducing the spread of infectious diseases like COVID-19 | 1.14% | 0.86% | 2.57% | 6.67% | 23.33% | 30.67% | 34.76% |
| Society's travel choices can contribute to reducing climate change | 1.90% | 0.95% | 2.19% | 6.57% | 25.62% | 31.05% | 31.71% |
| Society's travel choices can contribute to reducing the spread of infectious diseases like COVID-19 | 1.33% | 2.10% | 2.95% | 7.71% | 26.10% | 30.29% | 29.52% |

Table 4. Climate change & COVID-19 attitudes (importance, n=1,050) (Climate change = green fill; COVID-19 = blue fill)

| Question | Strongly disagree | Disagree | Somewhat disagree | Neither agree nor disagree | Somewhat agree | Agree | Strongly agree |
|---|---|---|---|---|---|---|---|
| It is important to **take actions** that reduce climate change | 1.62% | 0.95% | 1.81% | 7.52% | 19.81% | 30.67% | 37.62% |
| It is important to **take actions** that reduce the spread of infectious diseases like COVID-19 | 1.52% | 0.57% | 1.14% | 5.43% | 17.81% | 28.67% | 44.86% |
| It is important to **adopt travel habits** that reduce climate change | 1.81% | 1.52% | 2.38% | 11.14% | 24.19% | 30.10% | 28.86% |
| It is important to **adopt travel habits** that reduce the spread of infectious diseases like COVID-19 | 1.43% | 0.95% | 2.29% | 8.19% | 18.95% | 31.14% | 37.05% |

Table 5. Climate change & COVID-19 attitudes (ascription of responsibility, n=1,050) (Climate change = green fill; COVID-19 = blue fill)

| Question | Strongly disagree | Disagree | Somewhat disagree | Neither agree nor disagree | Somewhat agree | Agree | Strongly agree |
|---|---|---|---|---|---|---|---|
| It is **my responsibility** to alter my lifestyle and reduce the negative effects of climate change | 2.10% | 2.19% | 2.29% | 10.95% | 26.48% | 30.29% | 25.71% |
| It is **my responsibility** to alter my lifestyle to reduce the spread of infectious diseases like COVID-19 | 1.52% | 1.33% | 2.57% | 8.76% | 23.62% | 30.29% | 31.90% |
| It is **other people's responsibility**, not mine, to reduce the effects of climate change | 17.33% | 19.71% | 17.52% | 19.52% | 12.48% | 7.52% | 5.90% |
| It is **other people's responsibility**, not mine, to reduce the spread of infectious diseases like COVID-19 | 15.81% | 15.62% | 14.86% | 18.29% | 16.38% | 11.05% | 8.00% |
| It is **the Government's responsibility**, not mine, to reduce the effects of climate change | 8.95% | 10.38% | 12.19% | 18.76% | 23.24% | 13.90% | 12.57% |
| It is **the Government's responsibility**, not mine, to reduce the spread of infectious diseases like COVID-19 | 11.43% | 12.19% | 13.05% | 19.52% | 20.10% | 13.05% | 10.67% |
| It is the **responsibility of industry**, not mine, to reduce the effects of climate change | 8.67% | 11.33% | 12.48% | 18.57% | 22.57% | 13.62% | 12.76% |

Table 6. Willingness to adopt pro-environmental behaviours (n=1,050)

| Question | Extremely unlikely | Unlikely | Slightly unlikely | Neither likely nor unlikely | Slightly likely | Likely | Extremely likely |
|---|---|---|---|---|---|---|---|
| Recycling paper, glass and plastic | 1.43% | 0.57% | 0.76% | 3.71% | 6.48% | 22.67% | 64.38% |
| Reducing number of flights | 5.81% | 6.19% | 5.14% | 21.90% | 13.62% | 18.19% | 29.14% |
| Conserving energy | 1.90% | 1.52% | 2.67% | 9.81% | 21.43% | 34.95% | 27.71% |
| Purchasing only organic products | 15.33% | 15.14% | 15.43% | 20.48% | 17.52% | 10.48% | 5.62% |
| Switching from a petrol to an electric car | 11.33% | 8.10% | 10.19% | 21.05% | 16.38% | 18.29% | 14.67% |
| Becoming a member of an environmental group | 28.76% | 21.52% | 13.71% | 16.76% | 8.38% | 6.95% | 3.90% |
| Eating less meat | 14.19% | 11.24% | 13.33% | 13.71% | 19.33% | 14.48% | 13.71% |
| Using more public transportation | 10.19% | 11.62% | 11.14% | 15.43% | 17.81% | 16.95% | 16.86% |
| Walking or cycling more | 5.14% | 3.14% | 4.19% | 9.43% | 22.67% | 27.81% | 27.62% |

Table 7. Climate change & COVID-19 (social factors – social norms, social responsibility & perceived social capability) (n=1,050) (Climate change = green fill; COVID-19 = blue fill)

| Question | Strongly disagree | Disagree | Somewhat disagree | Neither agree nor disagree | Somewhat agree | Agree | Strongly agree |
|---|---|---|---|---|---|---|---|
| Most people who are important to me, are personally doing something to help reduce the risk of climate change | 4.38% | 7.33% | 8.29% | 26.86% | 30.57% | 14.48% | 8.10% |
| Most people who are important to me, are personally doing something to reduce their risk of contracting/spreading infectious diseases like COVID-19 | 1.71% | 1.81% | 4.29% | 12.19% | 32.67% | 24.00% | 23.33% |
| It is generally expected of me that I do my bit to help reduce the risk of climate change | 2.95% | 4.29% | 4.00% | 17.24% | 34.48% | 23.81% | 13.24% |
| It is generally expected of me that I do my bit to reduce the risk of contracting/spreading infectious diseases like COVID-19 | 1.24% | 1.14% | 2.29% | 8.95% | 30.38% | 27.24% | 28.76% |
| I feel that I should inspire people to take action to reduce the effects of climate change | 6.57% | 6.48% | 9.71% | 29.33% | 20.57% | 17.62% | 9.71% |
| I feel that I should inspire people to take action and reduce their risk of contracting/spreading infectious diseases like COVID-19 | 5.90% | 7.90% | 8.29% | 26.95% | 22.86% | 16.67% | 11.43% |
| I am often asked for advice by other people about ways to reduce the effects of climate change | 18.48% | 21.05% | 17.90% | 20.86% | 10.38% | 7.81% | 3.52% |
| I am often asked for advice by other people about ways to reduce their risk of contracting/spreading infectious diseases like COVID-19 | 14.10% | 16.86% | 15.90% | 24.67% | 12.57% | 10.29% | 5.62% |
| I feel that I am able to inspire people to take action and reduce the effects of climate change | 10.95% | 14.86% | 15.52% | 29.24% | 15.62% | 9.24% | 4.57% |
| I feel that I am able to inspire people to take action and reduce their risk of contracting/spreading infectious diseases like COVID-19 | 8.67% | 12.29% | 11.62% | 28.95% | 21.24% | 11.52% | 5.71% |

Table 8. Willingness to adopt behaviours that reduce the spread of infectious diseases (n=1,050)

| Question | Extremely unlikely | Unlikely | Slightly unlikely | Neither likely nor unlikely | Slightly likely | Likely | Extremely likely |
|---|---|---|---|---|---|---|---|
| Using public transport less | 6.29% | 7.90% | 8.10% | 19.24% | 11.62% | 20.95% | 25.90% |
| Walking or cycling more | 5.14% | 3.24% | 3.14% | 10.95% | 21.33% | 27.24% | 28.95% |
| Reducing my travel in general | 4.38% | 4.10% | 6.29% | 15.81% | 18.29% | 26.19% | 24.95% |
| Reducing number of flights | 5.24% | 4.67% | 5.05% | 18.86% | 11.71% | 18.19% | 36.29% |
| Following government guidelines | 2.38% | 1.33% | 2.29% | 7.52% | 12.10% | 30.00% | 44.38% |
| Working from home (if you had the chance) | 6.95% | 2.86% | 2.19% | 10.76% | 8.38% | 23.24% | 45.62% |
| Shopping online more | 4.57% | 4.00% | 4.38% | 16.48% | 17.24% | 25.14% | 28.19% |
| Reducing my social contacts | 5.52% | 7.05% | 8.19% | 15.81% | 16.48% | 24.38% | 22.57% |
| Volunteering for the NHS | 20.38% | 18.29% | 12.29% | 21.81% | 11.24% | 8.86% | 7.14% |